\documentclass[twocolumn, prl, 10pt]{revtex4-1}
\usepackage{graphicx, subfigure, amsfonts,amssymb,amsmath, array, gensymb,fontenc,color, lipsum}

\UseRawInputEncoding

\begin{document}

\title{{Imprinting tunable $\pi$-magnetism in graphene nanoribbons via edge extensions}}
\author{Michele Pizzochero}
\email{Electronic address: \url{mpizzochero@g.harvard.edu}}
\author{Efthimios Kaxiras}
\affiliation{School of Engineering and Applied Sciences, Harvard University, Cambridge, MA 02138, USA}

\date{\today}
\begin{abstract}

Magnetic carbon nanostructures are currently under scrutiny for a wide spectrum of applications. Here, we theoretically investigate armchair graphene nanoribbons patterned with asymmetric edge extensions consisting of laterally fused naphtho groups, as recently fabricated via on-surface synthesis. We show that an individual edge extension acts as a spin-$\frac{1}{2}$ center and develops a sizable spin-polarization of the conductance around the band edges. The Heisenberg exchange coupling between a pair of edge extensions is dictated by the position of the second naphtho group in the carbon backbone, thus enabling ferromagnetic, antiferromagnetic, or non-magnetic states. The periodic arrangement of edge extensions yields full spin-polarization at the band extrema, and the accompanying ferromagnetic ground state can be driven into non-magnetic or antiferromagnetic phases through external stimuli. Overall, our work reveals precise tunability of the $\pi$-magnetism in graphene nanoribbons induced by naphtho groups, thereby establishing these one-dimensional architectures as suitable platforms for logic spintronics.
\end{abstract}

\maketitle

 Although graphene exhibits a number of unique electronic properties \cite{Novo04}, including massless Dirac fermions and ballistic charge transport over microscopic length scales \cite{Novo05a, Neto09}, the vanishing band gap \cite{Neto09} and strong diamagnetic character \cite{Sepioni2010} have hindered its deployment in spin logic operations. In the quest of expanding the functionalities of graphene via quantum confinement effects \cite{Yaz13},  on-surface synthesis has proven successful to achieve graphene nanoribbons (GNRs) with desired atomic structures \cite{Cai10a, Nguy17}.  Within this bottom-up route, precursor organic molecules are self-assembled on a metal surface, yielding target GNRs that feature atomically precise edges \cite{Cai10a}. By tailor-making the initial precursor monomer, which encodes the topology of the final product, several GNRs with diverse geometries \cite{Cai10a, Ruffieux2016, Liu2015, Yano20} and widths \cite{Chen15a, Wang2017}  have been fabricated, thus stimulating the emergence of complex quantum phenomena \cite{Friedrich2020, Cao2017, Rizzo2018, Sun2020} and novel concepts for nanoscale devices \cite{Groning2018, Jacobse2017, Pizzochero2020, Ares07, Kang13, Kris2020}. 

In this vein, certain graphene nanoribbons have been envisaged as promising candidates for prospective logic spintronic components \cite{Yazyev2008, Son2006, Wimmer2008, Wang2008, Wang2009} by virtue of the favorable combination of weak spin-orbit and hyperfine interactions that ensure long spin lifetimes \cite{Avsar2020, Han2014} with fine-tunable band-gaps that confer switching capabilities \cite{Chen13, Son06a}. However, actual examples of intrinsic magnetism in these systems -- a key requirement in the context, for instance, of spin injection -- remain quite scarce. They are mainly restricted to either zigzag sites \cite{Ruffieux2016, Li2019}, end states \cite{Lawrence2020}, or heteroatom insertion \cite{Friedrich2020} in GNRs, invariably suffering from a limited control over the ensuing magnetism \cite{Yazyev2010, Radek18}. Desigllen
 extended carbon nanoarchitectures that offer versatile magnetic properties is an essential step toward the development of graphene-based switchable spintronics.  In this Letter, we investigate, from a theoretical point of view, armchair graphene nanoribbons functionalized with edge extensions, a class of structures that was recently experimentally realized via surface-assisted synthesis \cite{Sun2020m, Rizzo1597}. We demonstrate that a variety of magnetic states can be imprinted in graphene nanoribbons upon the incorporation of the edge extensions and subsequently manipulated through external stimuli, hence paving the way for all-carbon logic spintronics.

\begin{figure}[t]
  \centering
 \includegraphics[width=1\columnwidth]{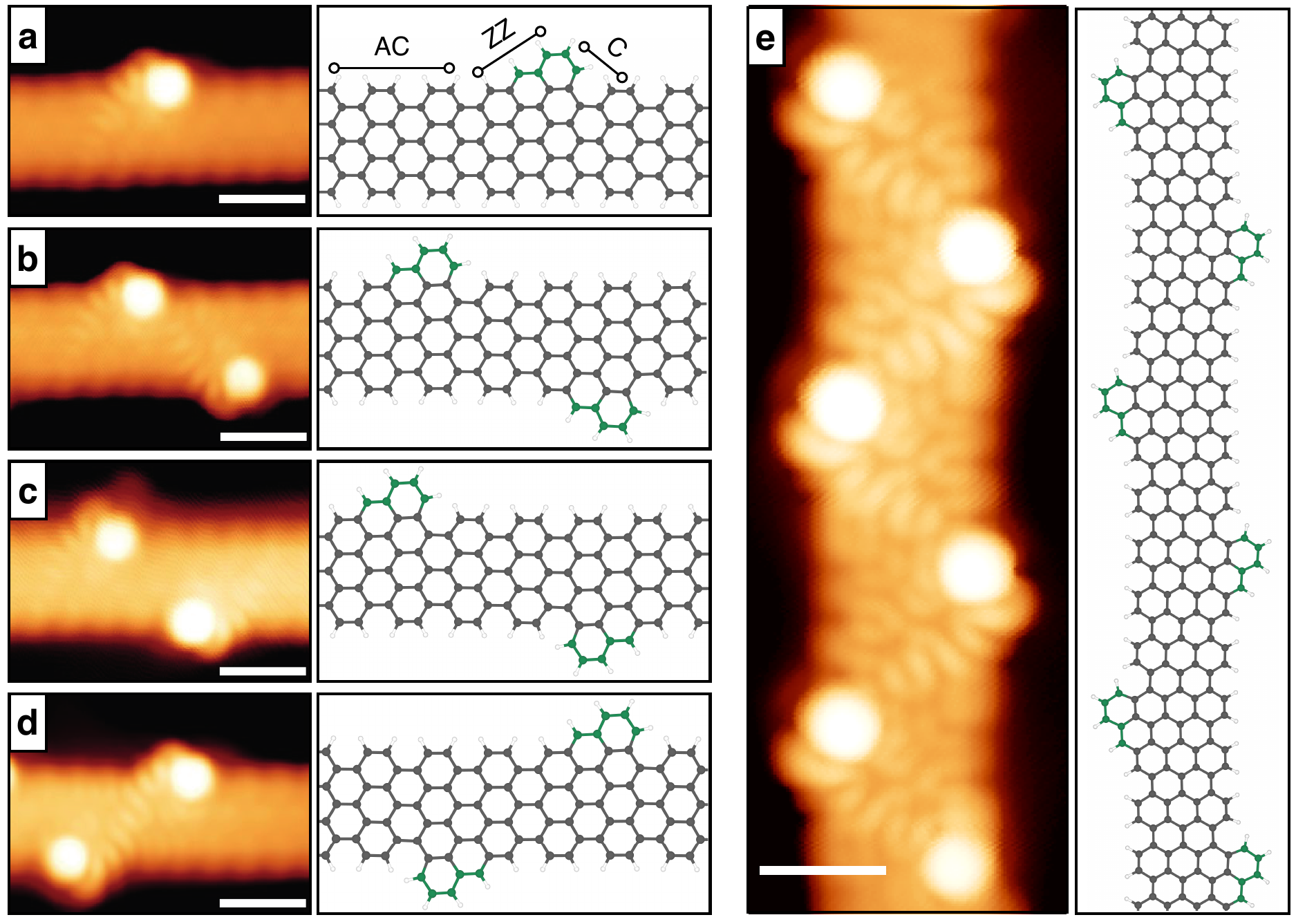}
  \caption{Constant-current STM images (left panels) and atomic structures (right panels) of 7-AGNR incorporating a (a) single edge-extension, double edge-extension in the (b) $D_1$, (c) $D_2$, and (d) $D_3$ dimer configuration, along with the (e) periodic sequence of such extensions, yielding sawtooth graphene nanoribbon. The edge-extension atoms are highlighted in green. In panel (a), the zigzag-edged (ZZ) extension-head and coved-edged (C) extension-tail are indicated.  Scale bars in STM images correspond to 1 nm. STM images are reproduced with permission from Ref.\ \cite{Sun2020m}, American Chemical Society.  \label{Fig1}}
\end{figure}

The nanostructures investigated in this work are 7-atom wide armchair graphene nanoribbons (7-AGNR) with edge extensions, as recently fabricated by Sun \emph{et al.}\ in Ref.\ \cite{Sun2020m}. Each edge extension consists of a naphtho group, that is, a pair of fused aromatic rings.  Five distinct configurations have been identified in experiments. An overview of their scanning-tunneling microscopy (STM) images is presented in Fig.\ \ref{Fig1}. In order to understand their electronic and magnetic structures, we combine first-principles and non-equilibrium Green's  function calculations. We rely on the generalized gradient approximation to density-functional theory devised by Perdew, Burke, and Ernzerhof \cite{PBE}, as implemented in the widely used \textsc{siesta} \cite{SIESTA} and \textsc{transiesta} \cite{TRANSIESTA} codes. Further computational details are given in Supplementary Note 1.

Fig.\ \ref{Fig1}(a) shows the atomic structure of a single edge-extension in 7-AGNR. The functionalization of the nanoribbon with the naphtho group yields an asymmetric extension that features a zigzag-edged \cite{Ruffieux2016} head and cove-edged \cite{Liu2015} tail (see Supplementary Fig.\ S1), expanding the armchair backbone by five carbon atoms. The addition of a second naphtho group in the vicinity of the first one occurs only in the three distinct dimer configurations labeled $D_1$ [Fig.\ \ref{Fig1}(b)], $D_2$ [Fig.\ \ref{Fig1}(c)],  and $D_3$ [Fig.\ \ref{Fig1}(d)]. Even though in all the three dimers the  7-AGNR scaffold is expanded by ten carbon atoms, the topology of the resulting nanoribbons differs in two important aspects. First, the relative distance between the pair of edge extensions, i.e., 1.50 nm, 1.64 nm, and 1.37 for the $D_1$, $D_2$, and $D_3$ configurations, respectively. Second, the mutual orientation (viz., the edge geometry of the extension-tail with respect to that of the adjacent extension-head) of the naphtho groups, namely, cove-edged to zigzag-edged in the $D_1$, cove-edged to cove-edged in the $D_2$, and zigzag-edged to zigzag-edged in the $D_3$ dimer configuration. The arrangement of the $D_1$ dimer in a periodic fashion leads to the array of edge extensions shown in Fig.\ \ref{Fig1}(e). The on-surface synthesis of this latter structural motif has been reported first in Ref.\ \citenum{Sun2020m} and later by Rizzo \emph{et al}.\ in Ref.\ \citenum{Rizzo1597}, who named it ``sawtooth" graphene nanoribbon.

 \begin{figure}[t]
  \centering
 \includegraphics[width=1\columnwidth]{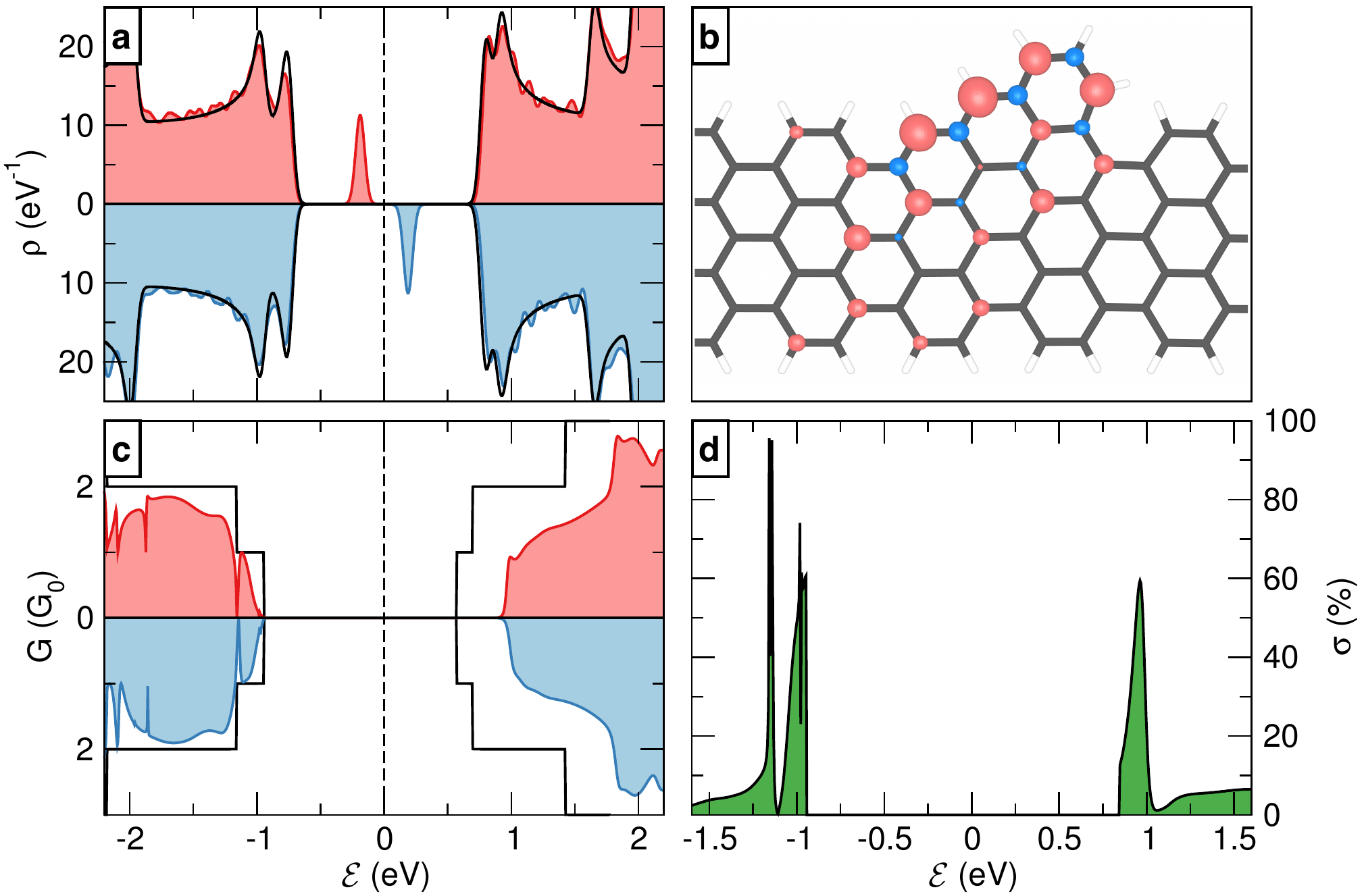}
  \caption{(a) Electronic density of states $\rho$ of 7-AGNR with (colored lines and areas) and without (black line) a single edge-extension, see Fig.\ \ref{Fig1}(a). (b) Spin density in the vicinity of the edge extension. (c) Conductance spectrum $G$ of 7-AGNR with (colored lines and areas) and without (black line) a single edge-extension. Red and blue colors in panels (b-d) indicate spin-majority and spin-minority channels, respectively. (d) Spin-polarization $\sigma$ of the conductance in 7-AGNR upon the addition of a single edge-extension.   \label{Fig2}}
\end{figure}

We begin our electronic structure investigation by considering the individual edge-extension in 7-AGNR shown Fig.\ \ref{Fig1}(a). The introduction of the naphtho group  breaks the sublattice symmetry, since the number of carbon atoms in the majority sublattice ($n\textsubscript{A}$) exceeds that of the minority sublattice ($n\textsubscript{B}$) by one unit.  This sublattice imbalance translates to a spin imbalance, sparking a magnetic moment of 1 $\mu\textsubscript{B}$. Such a magnetic solution lies 94 meV lower in energy than the non-magnetic one, hence indicating that the introduction of a single edge-extension drives a spin-$\frac{1}{2}$ ground state in otherwise non-magnetic 7-AGNR. These findings agree with Lieb's theorem for the repulsive Hubbard model of a bipartite lattice and a half-filled band \cite{Lieb1989}, which states that the magnetic ground state has spin $S = \frac{1}{2} |n\textsubscript{A} - n\textsubscript{B}|$. Although in the following we focus on 7-AGNR, the $S = \frac{1}{2}$ ground state induced by the naphtho group is robust against the width of the hosting armchair nanoribbon, as confirmed by our first-principles calculations.

In Fig.\ \ref{Fig2}(a), we show the electronic density of states to elucidate the origin of the magnetism induced by the edge extension in 7-AGNR. The addition of the naphtho group largely preserves the electronic structure of 7-AGNR, except for the emergence of a pair of in-gap spin-split states that are symmetrically located around the Fermi level and separated in energy by 0.38 eV. The magnetic moment that is encompassed in the singly occupied in-gap state primarily resides on the zigzag edge of the naphtho group and localizes on the majority sublattice in the vicinity of the extension, as displayed in Fig.\ \ref{Fig2}(b). Furthermore, the fused naphtho group largely disrupts the electronic transport across 7-AGNR, as we show in the conductance spectrum in Fig.\ \ref{Fig2}(c). This effect is particularly marked in the unoccupied states, where the conductance is found to be fully suppressed up to $\sim$0.5 eV above the conduction band maximum of bare 7-AGNR, effectively widening the transport gap.  Although detrimental to the charge transport, the introduction of the naphtho group and the ensuing magnetic moment cause a substantial spin-polarization [$\sigma(\cal{E})$] to arise.  As is customary, we quantify $\sigma(\cal{E})$ as

\begin{equation}
\sigma(\cal{E}) = \left|\frac{G_\uparrow(\cal{E}) - G_\downarrow(\cal{E})}{G_\uparrow(\cal{E}) + G_\downarrow(\cal{E})}\right| \%,
\end{equation}

\noindent
with $G_\uparrow(\cal{E})$ [$G_\downarrow(\cal{E})$] being the conductance spectrum of the majority [minority] spin channel. The result given in Fig.\ \ref{Fig2}(d) demonstrates that a spin-polarization of approximately 60\% is achieved in the vicinity of the band edges. This value can be largely increased (e.g., to more than 90\% at energy $\cal{E}$ $= -1.14$ eV) or decreased (e.g., to less than 2\% at $\cal{E}$ = 1.12 eV) through an appropriate modulation of the carrier density. Overall, these findings hint at potential spintronic applications of functionalized 7-AGNR even at a low density of edge extensions.

 \begin{figure}[t]
  \centering
 \includegraphics[width=1\columnwidth]{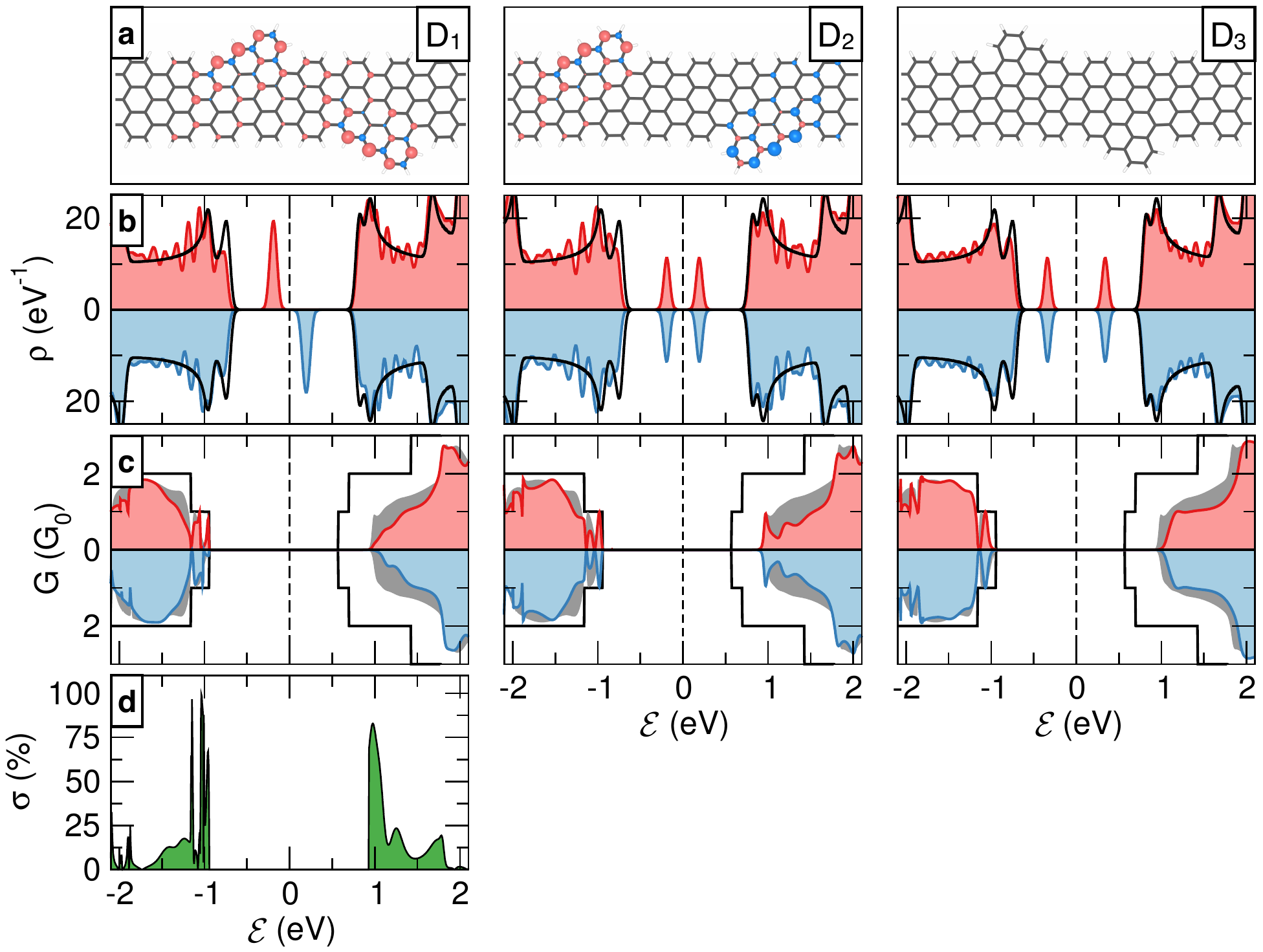}
  \caption{(a) Spin density in the vicinity of double edge-extensions in 7-AGNR in the $D_1$, $D_2$, and $D_3$ dimer configurations, see Fig.\ \ref{Fig1}(b-d). (b) Electronic density of states $\rho$ of 7-AGNR with a double (colored lines and areas) and without (black lines) edge-extension. (c) Conductance spectrum $G$ of 7-AGNR with a single (grey area), double (colored lines and areas), and without (black lines) edge-extension. Red and blue colors in panels (a-c) indicate spin-majority and spin-minority channels, respectively. (d) Spin-polarization $\sigma$ of the conductance in 7-AGNR upon the addition of a double edge-extension in the $D_1$ dimer configuration.  \label{Fig3}}
\end{figure}

 Next, we study pairs of edge extensions. The atomic structures are shown in Fig.\ \ref{Fig1}(b-d). A detailed analysis of their relative stability and aggregation tendency is provided in Supplementary Note 2.  Of these three structures obtained in experiments \cite{Sun2020m}, only that containing the $D_1$ dimer exhibits a sublattice imbalance, which in turn leads to a magnetic moment of 2 $\mu\textsubscript{B}$, that is, 1 $\mu\textsubscript{B}$ per naphtho group. This triplet state lies 182 meV lower in energy than the non-magnetic one. We assess the nature of the magnetic interactions between the two edge-extensions by estimating their Heisenberg exchange coupling, $J = \cal{E}\textsubscript{FM} - \cal{E}\textsubscript{AFM}$, where $\cal{E}\textsubscript{FM}$ and $\cal{E}\textsubscript{AFM}$ are the total energies of the investigated dimer in the ferromagnetic and antiferromagnetic configurations, respectively.
We obtain $J = -6$ meV, signaling a ferromagnetic coupling between the naphtho groups, as further supported by the spin density presented in Fig.\ \ref{Fig3}(a). These results establish a spin-$1$ ground state caused by the $D_1$ dimer, in line with the aforementioned Lieb's theorem \cite{Lieb1989}. The density of states given in Fig.\ \ref{Fig3}(b) reveals that, similarly to the single edge-extension case discussed above, the  magnetism associated with the $D_1$ dimer stems from a pair of (doubly occupied) spin-split states centered at the Fermi level, again 0.38 eV apart in energy.  As far as the charge transport is concerned, the introduction of a second naphtho group in the $D_1$ configuration further decreases the conductance of the functionalized 7-AGNR with respect to an individual edge-extension, see Fig. \ref{Fig3}(c). The doubling of the magnetic moment upon the formation of the
$D_1$ dimer enhances the spin-polarization around the band extrema, as we show in Fig.\ \ref{Fig3}(d). This is especially true at the conduction band edge, where $\sigma(\cal{E})$ attains a value of over 80\%.

Although the incorporation of two naphtho groups in the $D_2$ and $D_3$ dimer configurations preserves the sublattice symmetry in the hosting 7-AGNR, strikingly different magnetic ground states arise in the resulting nanoribbons. On the one hand, the addition of the $D_3$ dimer leaves the non-magnetic ground state of 7-AGNR unaffected [Fig.\ \ref{Fig3}(a)], being the triplet state 425 meV higher in energy. On the other hand, the addition of $D_2$ dimer develops a magnetic moment of 2 $\mu\textsubscript{B}$. For this latter configuration, however, we find a Heisenberg exchange coupling between the two naphtho groups $J = 4$ meV. The magnetic solution is 168 meV lower in energy than the non-magnetic one, thereby pointing to an antiferromagnetic (spin-$0$) ground state  [Fig.\ \ref{Fig3}(a)], in full compliance with  Lieb's theorem \cite{Lieb1989}. Thus, spin couplings are intertwined with the  positioning the naphtho groups in these AGNR. Despite these differences in the magnetic ground states of 7-AGNR upon the functionalization with the $D_2$ and $D_3$ dimers, similar changes in the electronic density of states are observed in Fig.\ \ref{Fig3}(b). In both cases, two pairs of spin-degenerate in-gap localized states emerge, whose separation in energy is 0.38 eV and 0.68 eV for the former and latter dimer, respectively. In Fig.\ \ref{Fig3}(c), we present the conductance spectra of 7-AGNR containing the $D_2$ and $D_3$ dimers, along with that of the single edge-extension. Compared to the single edge-extension, the formation of a second naphtho group in the $D_2$ ($D_3$) configuration retains (reduces)
the conductance at the band edges.

 \begin{figure*}[t]
  \centering
 \includegraphics[width=1.75\columnwidth]{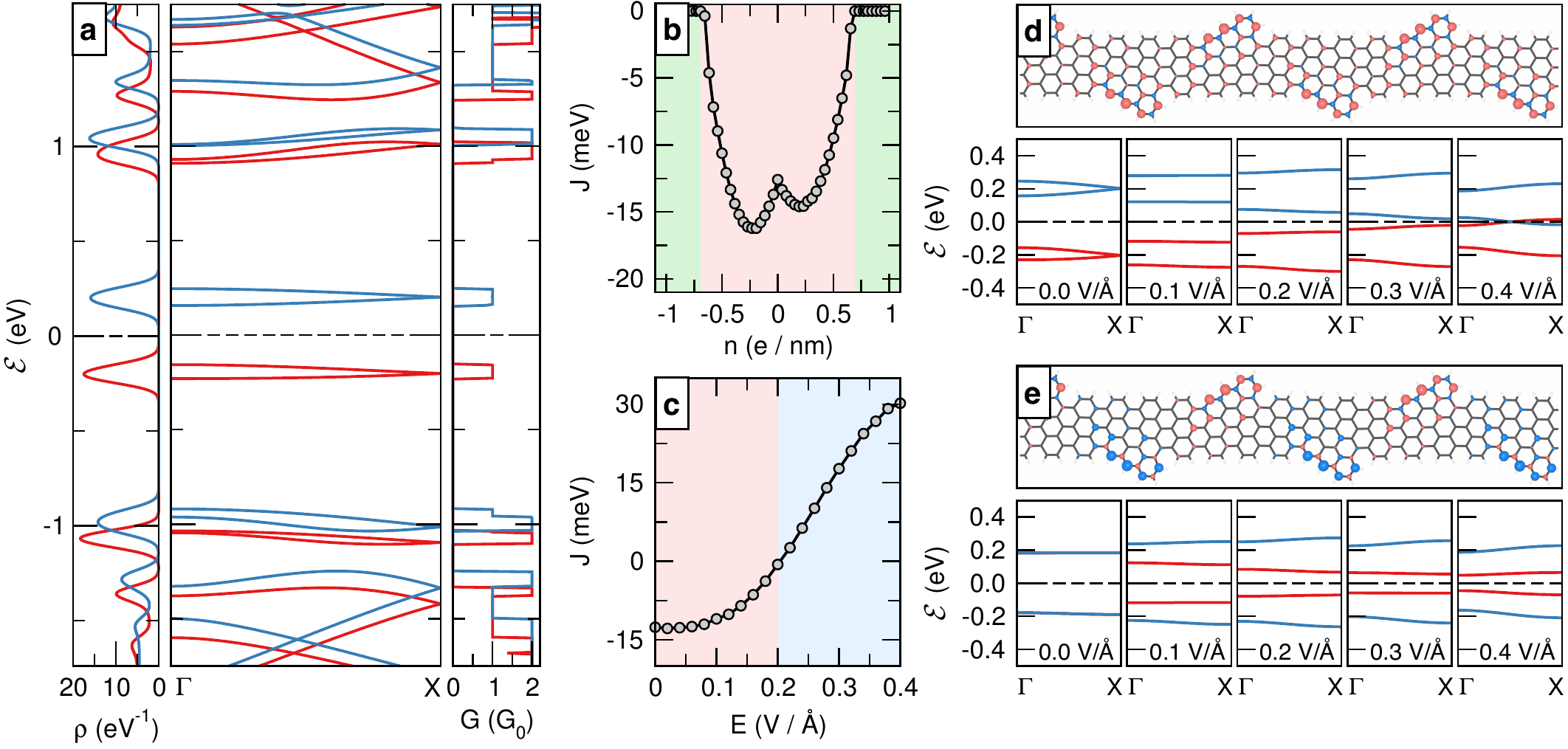}
  \caption{(a) Electronic structure of sGNR shown in Fig.\ \ref{Fig1}(e), comprising the electronic density of states $\rho$ (left panel), band structure (middle panel), and conductance spectrum $G$ (right panel). (b) Evolution of the Heisenberg exchange coupling $J$ with the charge doping $n$. Light red and light green backgrounds define the range of stability of the ferromagnetic and non-magnetic phases, respectively. (c) Evolution of the Heisenberg exchange coupling $J$ with the electric field $E$ applied in the tranverse direction. Light red and light blue backgrounds define the range of stability of the ferromagnetic and antiferromagnetic phases, respectively. (d) Spin density in the ferromagnetic phase of sGNR, along with the evolution of the band structure with increasing strength of the electric field. (e) Spin density in the antiferromagnetic phase of sGNR, along with the evolution of the band structure with increasing strength of the electric field. Red and blue colors in panels (a, d, e) indicate spin-majority and spin-minority channels, respectively. \label{Fig4}}
\end{figure*}

 Finally, we investigate the sawtooth graphene nanoribbon (sGNR) shown in Fig.\ \ref{Fig1}(e). As in the case of the $D_1$ dimer configuration, sGNR possesses a $S = 1$ ground state that lies 180 meV lower in energy than the non-magnetic one. The edge extensions are coupled by $J = -13$ meV, effectively acting as a ferromagnetic spin chain. This value is much larger than that reported for one-dimensional transition-metal chains \cite{Hirjibehedin1021}. The electronic structure of sGNR is given in Fig.\ \ref{Fig4}(a). The frontier bands feature a narrow bandwidth, which gives rise to a peaked density of states around the Fermi level. As the spin-majority channel encodes the valence band and the spin-minority the conduction band, a complete spin-polarization is achieved at both band edges, i.e.\ $\sigma(\cal{E})$ $= 100 \%$. This characteristic is insensitive to the width (see Supplementary Fig.\ S2) and edge passivation (see Supplementary Fig.\ S5) of the hosting nanoribbon, rendering sGNR a promising candidate for switchable spintronic components.

With the knowledge of the basic properties of sGNR at hand, we then explore its response to two external
perturbations, i.e., charge doping and electric field \cite{Jung2010}. Fig.\ \ref{Fig4}(b) presents the evolution of the Heisenberg exchange coupling in sGNR with both $n$-type and $p$-type doping. Two important effects take place. Firstly, we notice that a certain amount of extra charge appreciably strengthens $J$, which increases in magnitude from $-13$ meV at charge neutrality up to $-16$ meV ($-15$ meV) upon moderate electron (hole) doping. Second, we remark that carrier concentrations exceeding $|n| = 0.7$ $e$/nm (that is, two electrons or holes per unit cell) drive a ferromagnetic-to-non-magnetic phase transition, indicating that magnetism in sGNR can be switched on and off. The emergence of this doping-induced non-magnetic phase is width- and termination-independent (see Supplementary Fig.\ S3 and Fig.\ S5) and can be understood from the approximately symmetric nature of the density of states around the Fermi level shown in Fig.\ \ref{Fig4}(a). In analogy with previous observations in disordered graphene \cite{Pizzochero2015, Nair2013},  $n$-type doping populates the otherwise unoccupied electronic frontier energy level (the converse applies to $p$-type doping), decreasing and eventually quenching the magnetic moment. It is worth noticing that undesirable doping may prevent magnetism to be probed, e.g., if sGNR is placed or grown on a strongly interacting substrate. From the experimental point of view, this effect is of particular relevance in the framework of on-surface fabrication, where self-assembling of precursor molecules typically occurs on a gold surface, which in turn results in unintentional $p$-type doping of the as-synthesized samples \cite{Sun2020m, Rizzo1597}. 

Fig.\ \ref{Fig4}(c) shows the dependence of the Heisenberg exchange coupling in sawtooth graphene nanoribbon on an electric field ($E$) applied in the transverse direction. As the strength of the field exceeds the critical value of 0.2 V/{\AA}, the sign of $J$ reverts, thereby signaling a ferromagnetic-to-antiferromagnetic phase transition. The spin density pertaining to each of these magnetic states is displayed in Fig.\ \ref{Fig4}(d) and Fig.\ \ref{Fig4}(e), respectively. This field-controlled magnetic transition occurs irrespectively of the width of sGNR, albeit the critical value is dependent on the width and edge-passivation, see Supplementary Fig.\ S4 and Fig.\ S5. Furthermore, the magnitude of $J$ can be modulated by $E$ to a great extent (up to $30$ meV at $E = 0.4$ V/{\AA}), eventually surpassing the Landauer limit of minimum energy dissipation at room temperature, $k\textsubscript{B}T\ln(2) \simeq 18$ meV \cite{Landauer}.
The effect of the transverse electric field is not only restricted to the magnetism, but largely impacts the electronic properties as well. The evolution of the band structure of sGNR with $E$ in both ferro- and antiferro-magnetic phases is given in Fig.\ \ref{Fig4}(d-e). In the ferromagnetic phase, increasing the electric field reduces the gap, eventually reaching a band crossing at the Fermi level for $E = 0.4$ V/{\AA} \cite{Son2006}. Such a semiconductor-to-semimetal transition is accompanied by a halving of the magnetic moment. In the antiferromagnetic phase, on the other hand, the semiconducting character of sGNR is retained, yet the field-induced symmetry breaking enables to fine-tune the band-gap width for each spin channel separately, with the spin-majority channel being significantly narrower than the spin-minority one. Altogether, these results highlight pronounced magnetoelectric effects in sawtooth graphene nanoribbons \cite{Jung2010}.

In summary, we have found that an an individual edge-extension embedded in graphene nanoribbons gives rise to sublattice- and spin-imbalance, hence acting as a spin-$\frac{1}{2}$ center and causing a considerable spin-polarization of the conductance at the band edges.  Depending on the exact positioning of the second extension in the scaffold of the nanoribbon, ferromagnetic ($S = 1$), antiferromagnetic, or non-magnetic states ($S = 0$) arise. Upon the arrangement in a periodic array, these edge-extensions lead to a full spin-polarization at the band extrema. The accompanying ferromagnetic ground state can be guided into a non-magnetic or antiferromagnetic state through the application of charge doping or electric field, respectively. Our results show that unconventional $\pi$-magnetism in otherwise non-magnetic graphene nanoribbons can be induced and engineered \emph{\`a} \emph{la carte} by patterning the carbon skeleton with edge extensions and subsequent external stimuli. These nanoarchitectures possibly represent the ultimate limit of miniaturization for all-carbon spintronic devices.

\smallskip
M.P. gratefully acknowledges Q.\ Sun (Shanghai University), K.\ \v{C}er\c{n}evi\v{c}s (EPFL), G.\ Dellaferrera (IBM Research Europe), and Q.S.\ Wu (EPFL) for fruitful interactions. M.P. is financially supported by the Swiss National Science Foundation through the Early Postdoc.Mobility program (Grant No.\ P2ELP2-191706).


\begin{thebibliography}{46}%
\makeatletter
\providecommand \@ifxundefined [1]{%
 \@ifx{#1\undefined}
}%
\providecommand \@ifnum [1]{%
 \ifnum #1\expandafter \@firstoftwo
 \else \expandafter \@secondoftwo
 \fi
}%
\providecommand \@ifx [1]{%
 \ifx #1\expandafter \@firstoftwo
 \else \expandafter \@secondoftwo
 \fi
}%
\providecommand \natexlab [1]{#1}%
\providecommand \enquote  [1]{``#1''}%
\providecommand \bibnamefont  [1]{#1}%
\providecommand \bibfnamefont [1]{#1}%
\providecommand \citenamefont [1]{#1}%
\providecommand \href@noop [0]{\@secondoftwo}%
\providecommand \href [0]{\begingroup \@sanitize@url \@href}%
\providecommand \@href[1]{\@@startlink{#1}\@@href}%
\providecommand \@@href[1]{\endgroup#1\@@endlink}%
\providecommand \@sanitize@url [0]{\catcode `\\12\catcode `\$12\catcode
  `\&12\catcode `\#12\catcode `\^12\catcode `\_12\catcode `\%12\relax}%
\providecommand \@@startlink[1]{}%
\providecommand \@@endlink[0]{}%
\providecommand \url  [0]{\begingroup\@sanitize@url \@url }%
\providecommand \@url [1]{\endgroup\@href {#1}{\urlprefix }}%
\providecommand \urlprefix  [0]{URL }%
\providecommand \Eprint [0]{\href }%
\providecommand \doibase [0]{http://dx.doi.org/}%
\providecommand \selectlanguage [0]{\@gobble}%
\providecommand \bibinfo  [0]{\@secondoftwo}%
\providecommand \bibfield  [0]{\@secondoftwo}%
\providecommand \translation [1]{[#1]}%
\providecommand \BibitemOpen [0]{}%
\providecommand \bibitemStop [0]{}%
\providecommand \bibitemNoStop [0]{.\EOS\space}%
\providecommand \EOS [0]{\spacefactor3000\relax}%
\providecommand \BibitemShut  [1]{\csname bibitem#1\endcsname}%
\let\auto@bib@innerbib\@empty
\bibitem [{\citenamefont {Novoselov}\ \emph {et~al.}(2004)\citenamefont
  {Novoselov}, \citenamefont {Geim}, \citenamefont {Morozov}, \citenamefont
  {Jiang}, \citenamefont {Zhang}, \citenamefont {Dubonos}, \citenamefont
  {Grigorieva},\ and\ \citenamefont {Firsov}}]{Novo04}%
  \BibitemOpen
  \bibfield  {author} {\bibinfo {author} {\bibfnamefont {K.~S.}\ \bibnamefont
  {Novoselov}}, \bibinfo {author} {\bibfnamefont {A.~K.}\ \bibnamefont {Geim}},
  \bibinfo {author} {\bibfnamefont {S.~V.}\ \bibnamefont {Morozov}}, \bibinfo
  {author} {\bibfnamefont {D.}~\bibnamefont {Jiang}}, \bibinfo {author}
  {\bibfnamefont {Y.}~\bibnamefont {Zhang}}, \bibinfo {author} {\bibfnamefont
  {S.~V.}\ \bibnamefont {Dubonos}}, \bibinfo {author} {\bibfnamefont {I.~V.}\
  \bibnamefont {Grigorieva}}, \ and\ \bibinfo {author} {\bibfnamefont {A.~A.}\
  \bibnamefont {Firsov}},\ }\href {\doibase 10.1126/science.1102896} {\bibfield
   {journal} {\bibinfo  {journal} {Science}\ }\textbf {\bibinfo {volume}
  {306}},\ \bibinfo {pages} {666} (\bibinfo {year} {2004})}\BibitemShut
  {NoStop}%
\bibitem [{\citenamefont {Novoselov}\ \emph {et~al.}(2005)\citenamefont
  {Novoselov}, \citenamefont {Geim}, \citenamefont {Morozov}, \citenamefont
  {Jiang}, \citenamefont {Katsnelson}, \citenamefont {Grigorieva},
  \citenamefont {Dubonos},\ and\ \citenamefont {Firsov}}]{Novo05a}%
  \BibitemOpen
  \bibfield  {author} {\bibinfo {author} {\bibfnamefont {K.~S.}\ \bibnamefont
  {Novoselov}}, \bibinfo {author} {\bibfnamefont {A.~K.}\ \bibnamefont {Geim}},
  \bibinfo {author} {\bibfnamefont {S.~V.}\ \bibnamefont {Morozov}}, \bibinfo
  {author} {\bibfnamefont {D.}~\bibnamefont {Jiang}}, \bibinfo {author}
  {\bibfnamefont {M.~I.}\ \bibnamefont {Katsnelson}}, \bibinfo {author}
  {\bibfnamefont {I.~V.}\ \bibnamefont {Grigorieva}}, \bibinfo {author}
  {\bibfnamefont {S.~V.}\ \bibnamefont {Dubonos}}, \ and\ \bibinfo {author}
  {\bibfnamefont {A.~A.}\ \bibnamefont {Firsov}},\ }\href@noop {} {\bibfield
  {journal} {\bibinfo  {journal} {Nature}\ }\textbf {\bibinfo {volume} {438}},\
  \bibinfo {pages} {197} (\bibinfo {year} {2005})}\BibitemShut {NoStop}%
\bibitem [{\citenamefont {Castro~Neto}\ \emph {et~al.}(2009)\citenamefont
  {Castro~Neto}, \citenamefont {Guinea}, \citenamefont {Peres}, \citenamefont
  {Novoselov},\ and\ \citenamefont {Geim}}]{Neto09}%
  \BibitemOpen
  \bibfield  {author} {\bibinfo {author} {\bibfnamefont {A.~H.}\ \bibnamefont
  {Castro~Neto}}, \bibinfo {author} {\bibfnamefont {F.}~\bibnamefont {Guinea}},
  \bibinfo {author} {\bibfnamefont {N.~M.~R.}\ \bibnamefont {Peres}}, \bibinfo
  {author} {\bibfnamefont {K.~S.}\ \bibnamefont {Novoselov}}, \ and\ \bibinfo
  {author} {\bibfnamefont {A.~K.}\ \bibnamefont {Geim}},\ }\href@noop {}
  {\bibfield  {journal} {\bibinfo  {journal} {Rev. Mod. Phys.}\ }\textbf
  {\bibinfo {volume} {81}},\ \bibinfo {pages} {109} (\bibinfo {year}
  {2009})}\BibitemShut {NoStop}%
\bibitem [{\citenamefont {Sepioni}\ \emph {et~al.}(2010)\citenamefont
  {Sepioni}, \citenamefont {Nair}, \citenamefont {Rablen}, \citenamefont
  {Narayanan}, \citenamefont {Tuna}, \citenamefont {Winpenny}, \citenamefont
  {Geim},\ and\ \citenamefont {Grigorieva}}]{Sepioni2010}%
  \BibitemOpen
  \bibfield  {author} {\bibinfo {author} {\bibfnamefont {M.}~\bibnamefont
  {Sepioni}}, \bibinfo {author} {\bibfnamefont {R.~R.}\ \bibnamefont {Nair}},
  \bibinfo {author} {\bibfnamefont {S.}~\bibnamefont {Rablen}}, \bibinfo
  {author} {\bibfnamefont {J.}~\bibnamefont {Narayanan}}, \bibinfo {author}
  {\bibfnamefont {F.}~\bibnamefont {Tuna}}, \bibinfo {author} {\bibfnamefont
  {R.}~\bibnamefont {Winpenny}}, \bibinfo {author} {\bibfnamefont {A.~K.}\
  \bibnamefont {Geim}}, \ and\ \bibinfo {author} {\bibfnamefont {I.~V.}\
  \bibnamefont {Grigorieva}},\ }\href {\doibase 10.1103/PhysRevLett.105.207205}
  {\bibfield  {journal} {\bibinfo  {journal} {Phys. Rev. Lett.}\ }\textbf
  {\bibinfo {volume} {105}},\ \bibinfo {pages} {207205} (\bibinfo {year}
  {2010})}\BibitemShut {NoStop}%
\bibitem [{\citenamefont {Yazyev}(2013)}]{Yaz13}%
  \BibitemOpen
  \bibfield  {author} {\bibinfo {author} {\bibfnamefont {O.~V.}\ \bibnamefont
  {Yazyev}},\ }\href@noop {} {\bibfield  {journal} {\bibinfo  {journal} {Acc.
  Chem. Res.}\ }\textbf {\bibinfo {volume} {46}},\ \bibinfo {pages} {2319}
  (\bibinfo {year} {2013})}\BibitemShut {NoStop}%
\bibitem [{\citenamefont {Cai}\ \emph {et~al.}(2010)\citenamefont {Cai},
  \citenamefont {Ruffieux}, \citenamefont {Jaafar}, \citenamefont {Bieri},
  \citenamefont {Braun}, \citenamefont {Blankenburg}, \citenamefont {Muoth},
  \citenamefont {Seitsonen}, \citenamefont {Saleh}, \citenamefont {Feng},
  \citenamefont {Mullen},\ and\ \citenamefont {Fasel}}]{Cai10a}%
  \BibitemOpen
  \bibfield  {author} {\bibinfo {author} {\bibfnamefont {J.}~\bibnamefont
  {Cai}}, \bibinfo {author} {\bibfnamefont {P.}~\bibnamefont {Ruffieux}},
  \bibinfo {author} {\bibfnamefont {R.}~\bibnamefont {Jaafar}}, \bibinfo
  {author} {\bibfnamefont {M.}~\bibnamefont {Bieri}}, \bibinfo {author}
  {\bibfnamefont {T.}~\bibnamefont {Braun}}, \bibinfo {author} {\bibfnamefont
  {S.}~\bibnamefont {Blankenburg}}, \bibinfo {author} {\bibfnamefont
  {M.}~\bibnamefont {Muoth}}, \bibinfo {author} {\bibfnamefont {A.~P.}\
  \bibnamefont {Seitsonen}}, \bibinfo {author} {\bibfnamefont {M.}~\bibnamefont
  {Saleh}}, \bibinfo {author} {\bibfnamefont {X.}~\bibnamefont {Feng}},
  \bibinfo {author} {\bibfnamefont {K.}~\bibnamefont {Mullen}}, \ and\ \bibinfo
  {author} {\bibfnamefont {R.}~\bibnamefont {Fasel}},\ }\href@noop {}
  {\bibfield  {journal} {\bibinfo  {journal} {Nature}\ }\textbf {\bibinfo
  {volume} {466}},\ \bibinfo {pages} {470} (\bibinfo {year}
  {2010})}\BibitemShut {NoStop}%
\bibitem [{\citenamefont {Nguyen}\ \emph {et~al.}(2017)\citenamefont {Nguyen},
  \citenamefont {Tsai}, \citenamefont {Omrani}, \citenamefont {Marangoni},
  \citenamefont {Wu}, \citenamefont {Rizzo}, \citenamefont {Rodgers},
  \citenamefont {Cloke}, \citenamefont {Durr}, \citenamefont {Sakai},
  \citenamefont {Liou}, \citenamefont {Aikawa}, \citenamefont {Chelikowsky},
  \citenamefont {Louie}, \citenamefont {Fischer},\ and\ \citenamefont
  {Crommie}}]{Nguy17}%
  \BibitemOpen
  \bibfield  {author} {\bibinfo {author} {\bibfnamefont {G.~D.}\ \bibnamefont
  {Nguyen}}, \bibinfo {author} {\bibfnamefont {H.-Z.}\ \bibnamefont {Tsai}},
  \bibinfo {author} {\bibfnamefont {A.~A.}\ \bibnamefont {Omrani}}, \bibinfo
  {author} {\bibfnamefont {T.}~\bibnamefont {Marangoni}}, \bibinfo {author}
  {\bibfnamefont {M.}~\bibnamefont {Wu}}, \bibinfo {author} {\bibfnamefont
  {D.~J.}\ \bibnamefont {Rizzo}}, \bibinfo {author} {\bibfnamefont {G.~F.}\
  \bibnamefont {Rodgers}}, \bibinfo {author} {\bibfnamefont {R.~R.}\
  \bibnamefont {Cloke}}, \bibinfo {author} {\bibfnamefont {R.~A.}\ \bibnamefont
  {Durr}}, \bibinfo {author} {\bibfnamefont {Y.}~\bibnamefont {Sakai}},
  \bibinfo {author} {\bibfnamefont {F.}~\bibnamefont {Liou}}, \bibinfo {author}
  {\bibfnamefont {A.~S.}\ \bibnamefont {Aikawa}}, \bibinfo {author}
  {\bibfnamefont {J.~R.}\ \bibnamefont {Chelikowsky}}, \bibinfo {author}
  {\bibfnamefont {S.~G.}\ \bibnamefont {Louie}}, \bibinfo {author}
  {\bibfnamefont {F.~R.}\ \bibnamefont {Fischer}}, \ and\ \bibinfo {author}
  {\bibfnamefont {M.~F.}\ \bibnamefont {Crommie}},\ }\href@noop {} {\bibfield
  {journal} {\bibinfo  {journal} {Nat. Nanotechnol.}\ }\textbf {\bibinfo
  {volume} {12}},\ \bibinfo {pages} {1077} (\bibinfo {year}
  {2017})}\BibitemShut {NoStop}%
\bibitem [{\citenamefont {Ruffieux}\ \emph {et~al.}(2016)\citenamefont
  {Ruffieux}, \citenamefont {Wang}, \citenamefont {Yang}, \citenamefont
  {S{\'a}nchez-S{\'a}nchez}, \citenamefont {Liu}, \citenamefont {Dienel},
  \citenamefont {Talirz}, \citenamefont {Shinde}, \citenamefont {Pignedoli},
  \citenamefont {Passerone}, \citenamefont {Dumslaff}, \citenamefont {Feng},
  \citenamefont {M{\"u}llen},\ and\ \citenamefont {Fasel}}]{Ruffieux2016}%
  \BibitemOpen
  \bibfield  {author} {\bibinfo {author} {\bibfnamefont {P.}~\bibnamefont
  {Ruffieux}}, \bibinfo {author} {\bibfnamefont {S.}~\bibnamefont {Wang}},
  \bibinfo {author} {\bibfnamefont {B.}~\bibnamefont {Yang}}, \bibinfo {author}
  {\bibfnamefont {C.}~\bibnamefont {S{\'a}nchez-S{\'a}nchez}}, \bibinfo
  {author} {\bibfnamefont {J.}~\bibnamefont {Liu}}, \bibinfo {author}
  {\bibfnamefont {T.}~\bibnamefont {Dienel}}, \bibinfo {author} {\bibfnamefont
  {L.}~\bibnamefont {Talirz}}, \bibinfo {author} {\bibfnamefont
  {P.}~\bibnamefont {Shinde}}, \bibinfo {author} {\bibfnamefont {C.~A.}\
  \bibnamefont {Pignedoli}}, \bibinfo {author} {\bibfnamefont {D.}~\bibnamefont
  {Passerone}}, \bibinfo {author} {\bibfnamefont {T.}~\bibnamefont {Dumslaff}},
  \bibinfo {author} {\bibfnamefont {X.}~\bibnamefont {Feng}}, \bibinfo {author}
  {\bibfnamefont {K.}~\bibnamefont {M{\"u}llen}}, \ and\ \bibinfo {author}
  {\bibfnamefont {R.}~\bibnamefont {Fasel}},\ }\href {\doibase
  10.1038/nature17151} {\bibfield  {journal} {\bibinfo  {journal} {Nature}\
  }\textbf {\bibinfo {volume} {531}},\ \bibinfo {pages} {489} (\bibinfo {year}
  {2016})}\BibitemShut {NoStop}%
\bibitem [{\citenamefont {Liu}\ \emph {et~al.}(2015)\citenamefont {Liu},
  \citenamefont {Li}, \citenamefont {Tan}, \citenamefont {Giannakopoulos},
  \citenamefont {S{\'a}nchez-S{\'a}nchez}, \citenamefont {Beljonne},
  \citenamefont {Ruffieux}, \citenamefont {Fasel}, \citenamefont {Feng},\ and\
  \citenamefont {M\"{u}llen}}]{Liu2015}%
  \BibitemOpen
  \bibfield  {author} {\bibinfo {author} {\bibfnamefont {J.}~\bibnamefont
  {Liu}}, \bibinfo {author} {\bibfnamefont {B.-W.}\ \bibnamefont {Li}},
  \bibinfo {author} {\bibfnamefont {Y.-Z.}\ \bibnamefont {Tan}}, \bibinfo
  {author} {\bibfnamefont {A.}~\bibnamefont {Giannakopoulos}}, \bibinfo
  {author} {\bibfnamefont {C.}~\bibnamefont {S{\'a}nchez-S{\'a}nchez}},
  \bibinfo {author} {\bibfnamefont {D.}~\bibnamefont {Beljonne}}, \bibinfo
  {author} {\bibfnamefont {P.}~\bibnamefont {Ruffieux}}, \bibinfo {author}
  {\bibfnamefont {R.}~\bibnamefont {Fasel}}, \bibinfo {author} {\bibfnamefont
  {X.}~\bibnamefont {Feng}}, \ and\ \bibinfo {author} {\bibfnamefont
  {K.}~\bibnamefont {M\"{u}llen}},\ }\href {\doibase 10.1021/jacs.5b03017}
  {\bibfield  {journal} {\bibinfo  {journal} {J. Am. Chem. Soc.}\ }\textbf
  {\bibinfo {volume} {137}},\ \bibinfo {pages} {6097} (\bibinfo {year}
  {2015})}\BibitemShut {NoStop}%
\bibitem [{\citenamefont {Yano}\ \emph {et~al.}(2020)\citenamefont {Yano},
  \citenamefont {Mitoma}, \citenamefont {Ito},\ and\ \citenamefont
  {Itami}}]{Yano20}%
  \BibitemOpen
  \bibfield  {author} {\bibinfo {author} {\bibfnamefont {Y.}~\bibnamefont
  {Yano}}, \bibinfo {author} {\bibfnamefont {N.}~\bibnamefont {Mitoma}},
  \bibinfo {author} {\bibfnamefont {H.}~\bibnamefont {Ito}}, \ and\ \bibinfo
  {author} {\bibfnamefont {K.}~\bibnamefont {Itami}},\ }\href@noop {}
  {\bibfield  {journal} {\bibinfo  {journal} {J. Org. Chem.}\ }\textbf
  {\bibinfo {volume} {85}},\ \bibinfo {pages} {4} (\bibinfo {year}
  {2020})}\BibitemShut {NoStop}%
\bibitem [{\citenamefont {Chen}\ \emph {et~al.}(2015)\citenamefont {Chen},
  \citenamefont {Cao}, \citenamefont {Chen}, \citenamefont {Pedramrazi},
  \citenamefont {Haberer}, \citenamefont {de~Oteyza}, \citenamefont {Fischer},
  \citenamefont {Louie},\ and\ \citenamefont {Crommie}}]{Chen15a}%
  \BibitemOpen
  \bibfield  {author} {\bibinfo {author} {\bibfnamefont {Y.-C.}\ \bibnamefont
  {Chen}}, \bibinfo {author} {\bibfnamefont {T.}~\bibnamefont {Cao}}, \bibinfo
  {author} {\bibfnamefont {C.}~\bibnamefont {Chen}}, \bibinfo {author}
  {\bibfnamefont {Z.}~\bibnamefont {Pedramrazi}}, \bibinfo {author}
  {\bibfnamefont {D.}~\bibnamefont {Haberer}}, \bibinfo {author} {\bibfnamefont
  {D.~G.}\ \bibnamefont {de~Oteyza}}, \bibinfo {author} {\bibfnamefont {F.~R.}\
  \bibnamefont {Fischer}}, \bibinfo {author} {\bibfnamefont {S.~G.}\
  \bibnamefont {Louie}}, \ and\ \bibinfo {author} {\bibfnamefont {M.~F.}\
  \bibnamefont {Crommie}},\ }\href@noop {} {\bibfield  {journal} {\bibinfo
  {journal} {Nat. Nanotechnol.}\ }\textbf {\bibinfo {volume} {10}},\ \bibinfo
  {pages} {156} (\bibinfo {year} {2015})}\BibitemShut {NoStop}%
\bibitem [{\citenamefont {Wang}\ \emph {et~al.}(2017)\citenamefont {Wang},
  \citenamefont {Kharche}, \citenamefont {Costa~Girão}, \citenamefont {Feng},
  \citenamefont {Müllen}, \citenamefont {Meunier}, \citenamefont {Fasel},\
  and\ \citenamefont {Ruffieux}}]{Wang2017}%
  \BibitemOpen
  \bibfield  {author} {\bibinfo {author} {\bibfnamefont {S.}~\bibnamefont
  {Wang}}, \bibinfo {author} {\bibfnamefont {N.}~\bibnamefont {Kharche}},
  \bibinfo {author} {\bibfnamefont {E.}~\bibnamefont {Costa~Girão}}, \bibinfo
  {author} {\bibfnamefont {X.}~\bibnamefont {Feng}}, \bibinfo {author}
  {\bibfnamefont {K.}~\bibnamefont {Müllen}}, \bibinfo {author} {\bibfnamefont
  {V.}~\bibnamefont {Meunier}}, \bibinfo {author} {\bibfnamefont
  {R.}~\bibnamefont {Fasel}}, \ and\ \bibinfo {author} {\bibfnamefont
  {P.}~\bibnamefont {Ruffieux}},\ }\href {\doibase
  10.1021/acs.nanolett.7b01244} {\bibfield  {journal} {\bibinfo  {journal}
  {Nano Lett.}\ }\textbf {\bibinfo {volume} {17}},\ \bibinfo {pages} {4277}
  (\bibinfo {year} {2017})}\BibitemShut {NoStop}%
\bibitem [{\citenamefont {Friedrich}\ \emph {et~al.}(2020)\citenamefont
  {Friedrich}, \citenamefont {Brandimarte}, \citenamefont {Li}, \citenamefont
  {Saito}, \citenamefont {Yamaguchi}, \citenamefont {Pozo}, \citenamefont
  {Pe\~na}, \citenamefont {Frederiksen}, \citenamefont {Garcia-Lekue},
  \citenamefont {S{\'a}nchez-Portal},\ and\ \citenamefont
  {Pascual}}]{Friedrich2020}%
  \BibitemOpen
  \bibfield  {author} {\bibinfo {author} {\bibfnamefont {N.}~\bibnamefont
  {Friedrich}}, \bibinfo {author} {\bibfnamefont {P.}~\bibnamefont
  {Brandimarte}}, \bibinfo {author} {\bibfnamefont {J.}~\bibnamefont {Li}},
  \bibinfo {author} {\bibfnamefont {S.}~\bibnamefont {Saito}}, \bibinfo
  {author} {\bibfnamefont {S.}~\bibnamefont {Yamaguchi}}, \bibinfo {author}
  {\bibfnamefont {I.}~\bibnamefont {Pozo}}, \bibinfo {author} {\bibfnamefont
  {D.}~\bibnamefont {Pe\~na}}, \bibinfo {author} {\bibfnamefont
  {T.}~\bibnamefont {Frederiksen}}, \bibinfo {author} {\bibfnamefont
  {A.}~\bibnamefont {Garcia-Lekue}}, \bibinfo {author} {\bibfnamefont
  {D.}~\bibnamefont {S{\'a}nchez-Portal}}, \ and\ \bibinfo {author}
  {\bibfnamefont {J.~I.}\ \bibnamefont {Pascual}},\ }\href@noop {} {\bibfield
  {journal} {\bibinfo  {journal} {Phys. Rev. Lett.}\ }\textbf {\bibinfo
  {volume} {125}},\ \bibinfo {pages} {146801} (\bibinfo {year}
  {2020})}\BibitemShut {NoStop}%
\bibitem [{\citenamefont {Cao}\ \emph {et~al.}(2017)\citenamefont {Cao},
  \citenamefont {Zhao},\ and\ \citenamefont {Louie}}]{Cao2017}%
  \BibitemOpen
  \bibfield  {author} {\bibinfo {author} {\bibfnamefont {T.}~\bibnamefont
  {Cao}}, \bibinfo {author} {\bibfnamefont {F.}~\bibnamefont {Zhao}}, \ and\
  \bibinfo {author} {\bibfnamefont {S.~G.}\ \bibnamefont {Louie}},\ }\href
  {\doibase 10.1103/PhysRevLett.119.076401} {\bibfield  {journal} {\bibinfo
  {journal} {Phys. Rev. Lett.}\ }\textbf {\bibinfo {volume} {119}},\ \bibinfo
  {pages} {076401} (\bibinfo {year} {2017})}\BibitemShut {NoStop}%
\bibitem [{\citenamefont {Rizzo}\ \emph {et~al.}(2018)\citenamefont {Rizzo},
  \citenamefont {Veber}, \citenamefont {Cao}, \citenamefont {Bronner},
  \citenamefont {Chen}, \citenamefont {Zhao}, \citenamefont {Rodriguez},
  \citenamefont {Louie}, \citenamefont {Crommie},\ and\ \citenamefont
  {Fischer}}]{Rizzo2018}%
  \BibitemOpen
  \bibfield  {author} {\bibinfo {author} {\bibfnamefont {D.~J.}\ \bibnamefont
  {Rizzo}}, \bibinfo {author} {\bibfnamefont {G.}~\bibnamefont {Veber}},
  \bibinfo {author} {\bibfnamefont {T.}~\bibnamefont {Cao}}, \bibinfo {author}
  {\bibfnamefont {C.}~\bibnamefont {Bronner}}, \bibinfo {author} {\bibfnamefont
  {T.}~\bibnamefont {Chen}}, \bibinfo {author} {\bibfnamefont {F.}~\bibnamefont
  {Zhao}}, \bibinfo {author} {\bibfnamefont {H.}~\bibnamefont {Rodriguez}},
  \bibinfo {author} {\bibfnamefont {S.~G.}\ \bibnamefont {Louie}}, \bibinfo
  {author} {\bibfnamefont {M.~F.}\ \bibnamefont {Crommie}}, \ and\ \bibinfo
  {author} {\bibfnamefont {F.~R.}\ \bibnamefont {Fischer}},\ }\href {\doibase
  10.1038/s41586-018-0376-8} {\bibfield  {journal} {\bibinfo  {journal}
  {Nature}\ }\textbf {\bibinfo {volume} {560}},\ \bibinfo {pages} {204}
  (\bibinfo {year} {2018})}\BibitemShut {NoStop}%
\bibitem [{\citenamefont {Sun}\ \emph {et~al.}(2020{\natexlab{a}})\citenamefont
  {Sun}, \citenamefont {Gr{\"o}ning}, \citenamefont {Overbeck}, \citenamefont
  {Braun}, \citenamefont {Perrin}, \citenamefont {Borin~Barin}, \citenamefont
  {El~Abbassi}, \citenamefont {Eimre}, \citenamefont {Ditler}, \citenamefont
  {Daniels}, \citenamefont {Meunier}, \citenamefont {Pignedoli}, \citenamefont
  {Calame}, \citenamefont {Fasel},\ and\ \citenamefont {Ruffieux}}]{Sun2020}%
  \BibitemOpen
  \bibfield  {author} {\bibinfo {author} {\bibfnamefont {Q.}~\bibnamefont
  {Sun}}, \bibinfo {author} {\bibfnamefont {O.}~\bibnamefont {Gr{\"o}ning}},
  \bibinfo {author} {\bibfnamefont {J.}~\bibnamefont {Overbeck}}, \bibinfo
  {author} {\bibfnamefont {O.}~\bibnamefont {Braun}}, \bibinfo {author}
  {\bibfnamefont {M.~L.}\ \bibnamefont {Perrin}}, \bibinfo {author}
  {\bibfnamefont {G.}~\bibnamefont {Borin~Barin}}, \bibinfo {author}
  {\bibfnamefont {M.}~\bibnamefont {El~Abbassi}}, \bibinfo {author}
  {\bibfnamefont {K.}~\bibnamefont {Eimre}}, \bibinfo {author} {\bibfnamefont
  {E.}~\bibnamefont {Ditler}}, \bibinfo {author} {\bibfnamefont
  {C.}~\bibnamefont {Daniels}}, \bibinfo {author} {\bibfnamefont
  {V.}~\bibnamefont {Meunier}}, \bibinfo {author} {\bibfnamefont {C.~A.}\
  \bibnamefont {Pignedoli}}, \bibinfo {author} {\bibfnamefont {M.}~\bibnamefont
  {Calame}}, \bibinfo {author} {\bibfnamefont {R.}~\bibnamefont {Fasel}}, \
  and\ \bibinfo {author} {\bibfnamefont {P.}~\bibnamefont {Ruffieux}},\
  }\href@noop {} {\bibfield  {journal} {\bibinfo  {journal} {Adv. Mater.}\
  }\textbf {\bibinfo {volume} {32}},\ \bibinfo {pages} {1906054} (\bibinfo
  {year} {2020}{\natexlab{a}})}\BibitemShut {NoStop}%
\bibitem [{\citenamefont {Gr{\"o}ning}\ \emph {et~al.}(2018)\citenamefont
  {Gr{\"o}ning}, \citenamefont {Wang}, \citenamefont {Yao}, \citenamefont
  {Pignedoli}, \citenamefont {Borin~Barin}, \citenamefont {Daniels},
  \citenamefont {Cupo}, \citenamefont {Meunier}, \citenamefont {Feng},
  \citenamefont {Narita}, \citenamefont {M\"ollen}, \citenamefont {Ruffieux},\
  and\ \citenamefont {Fasel}}]{Groning2018}%
  \BibitemOpen
  \bibfield  {author} {\bibinfo {author} {\bibfnamefont {O.}~\bibnamefont
  {Gr{\"o}ning}}, \bibinfo {author} {\bibfnamefont {S.}~\bibnamefont {Wang}},
  \bibinfo {author} {\bibfnamefont {X.}~\bibnamefont {Yao}}, \bibinfo {author}
  {\bibfnamefont {C.~A.}\ \bibnamefont {Pignedoli}}, \bibinfo {author}
  {\bibfnamefont {G.}~\bibnamefont {Borin~Barin}}, \bibinfo {author}
  {\bibfnamefont {C.}~\bibnamefont {Daniels}}, \bibinfo {author} {\bibfnamefont
  {A.}~\bibnamefont {Cupo}}, \bibinfo {author} {\bibfnamefont {V.}~\bibnamefont
  {Meunier}}, \bibinfo {author} {\bibfnamefont {X.}~\bibnamefont {Feng}},
  \bibinfo {author} {\bibfnamefont {A.}~\bibnamefont {Narita}}, \bibinfo
  {author} {\bibfnamefont {K.}~\bibnamefont {M\"ollen}}, \bibinfo {author}
  {\bibfnamefont {P.}~\bibnamefont {Ruffieux}}, \ and\ \bibinfo {author}
  {\bibfnamefont {R.}~\bibnamefont {Fasel}},\ }\href@noop {} {\bibfield
  {journal} {\bibinfo  {journal} {Nature}\ }\textbf {\bibinfo {volume} {560}},\
  \bibinfo {pages} {209} (\bibinfo {year} {2018})}\BibitemShut {NoStop}%
\bibitem [{\citenamefont {Jacobse}\ \emph {et~al.}(2017)\citenamefont
  {Jacobse}, \citenamefont {Kimouche}, \citenamefont {Gebraad}, \citenamefont
  {Ervasti}, \citenamefont {Thijssen}, \citenamefont {Liljeroth},\ and\
  \citenamefont {Swart}}]{Jacobse2017}%
  \BibitemOpen
  \bibfield  {author} {\bibinfo {author} {\bibfnamefont {P.~H.}\ \bibnamefont
  {Jacobse}}, \bibinfo {author} {\bibfnamefont {A.}~\bibnamefont {Kimouche}},
  \bibinfo {author} {\bibfnamefont {T.}~\bibnamefont {Gebraad}}, \bibinfo
  {author} {\bibfnamefont {M.~M.}\ \bibnamefont {Ervasti}}, \bibinfo {author}
  {\bibfnamefont {J.~M.}\ \bibnamefont {Thijssen}}, \bibinfo {author}
  {\bibfnamefont {P.}~\bibnamefont {Liljeroth}}, \ and\ \bibinfo {author}
  {\bibfnamefont {I.}~\bibnamefont {Swart}},\ }\href {\doibase
  10.1038/s41467-017-00195-2} {\bibfield  {journal} {\bibinfo  {journal} {Nat.
  Commun.}\ }\textbf {\bibinfo {volume} {8}},\ \bibinfo {pages} {119} (\bibinfo
  {year} {2017})}\BibitemShut {NoStop}%
\bibitem [{\citenamefont {\v{C}er\c{n}evi\v{c}s}\ \emph
  {et~al.}(2020{\natexlab{a}})\citenamefont {\v{C}er\c{n}evi\v{c}s},
  \citenamefont {Yazyev},\ and\ \citenamefont {Pizzochero}}]{Pizzochero2020}%
  \BibitemOpen
  \bibfield  {author} {\bibinfo {author} {\bibfnamefont {K.}~\bibnamefont
  {\v{C}er\c{n}evi\v{c}s}}, \bibinfo {author} {\bibfnamefont {O.~V.}\
  \bibnamefont {Yazyev}}, \ and\ \bibinfo {author} {\bibfnamefont
  {M.}~\bibnamefont {Pizzochero}},\ }\href {\doibase
  10.1103/PhysRevB.102.201406} {\bibfield  {journal} {\bibinfo  {journal}
  {Phys. Rev. B}\ }\textbf {\bibinfo {volume} {102}},\ \bibinfo {pages}
  {201406} (\bibinfo {year} {2020}{\natexlab{a}})}\BibitemShut {NoStop}%
\bibitem [{\citenamefont {Areshkin}\ and\ \citenamefont
  {White}(2007)}]{Ares07}%
  \BibitemOpen
  \bibfield  {author} {\bibinfo {author} {\bibfnamefont {D.~A.}\ \bibnamefont
  {Areshkin}}\ and\ \bibinfo {author} {\bibfnamefont {C.~T.}\ \bibnamefont
  {White}},\ }\href@noop {} {\bibfield  {journal} {\bibinfo  {journal} {Nano
  Lett.}\ }\textbf {\bibinfo {volume} {7}},\ \bibinfo {pages} {3253} (\bibinfo
  {year} {2007})}\BibitemShut {NoStop}%
\bibitem [{\citenamefont {Kang}\ \emph {et~al.}(2013)\citenamefont {Kang},
  \citenamefont {Sarkar}, \citenamefont {Khatami},\ and\ \citenamefont
  {Banerjee}}]{Kang13}%
  \BibitemOpen
  \bibfield  {author} {\bibinfo {author} {\bibfnamefont {J.}~\bibnamefont
  {Kang}}, \bibinfo {author} {\bibfnamefont {D.}~\bibnamefont {Sarkar}},
  \bibinfo {author} {\bibfnamefont {Y.}~\bibnamefont {Khatami}}, \ and\
  \bibinfo {author} {\bibfnamefont {K.}~\bibnamefont {Banerjee}},\ }\href@noop
  {} {\bibfield  {journal} {\bibinfo  {journal} {Appl. Phys. Lett.}\ }\textbf
  {\bibinfo {volume} {103}},\ \bibinfo {pages} {083113} (\bibinfo {year}
  {2013})}\BibitemShut {NoStop}%
\bibitem [{\citenamefont {\v{C}er\c{n}evi\v{c}s}\ \emph
  {et~al.}(2020{\natexlab{b}})\citenamefont {\v{C}er\c{n}evi\v{c}s},
  \citenamefont {Pizzochero},\ and\ \citenamefont {Yazyev}}]{Kris2020}%
  \BibitemOpen
  \bibfield  {author} {\bibinfo {author} {\bibfnamefont {K.}~\bibnamefont
  {\v{C}er\c{n}evi\v{c}s}}, \bibinfo {author} {\bibfnamefont {M.}~\bibnamefont
  {Pizzochero}}, \ and\ \bibinfo {author} {\bibfnamefont {O.~V.}\ \bibnamefont
  {Yazyev}},\ }\href {\doibase 10.1140/epjp/s13360-020-00696-y} {\bibfield
  {journal} {\bibinfo  {journal} {Eur. Phys. J. Plus}\ }\textbf {\bibinfo
  {volume} {135}},\ \bibinfo {pages} {681} (\bibinfo {year}
  {2020}{\natexlab{b}})}\BibitemShut {NoStop}%
\bibitem [{\citenamefont {Yazyev}\ and\ \citenamefont
  {Katsnelson}(2008)}]{Yazyev2008}%
  \BibitemOpen
  \bibfield  {author} {\bibinfo {author} {\bibfnamefont {O.~V.}\ \bibnamefont
  {Yazyev}}\ and\ \bibinfo {author} {\bibfnamefont {M.~I.}\ \bibnamefont
  {Katsnelson}},\ }\href {\doibase 10.1103/PhysRevLett.100.047209} {\bibfield
  {journal} {\bibinfo  {journal} {Phys. Rev. Lett.}\ }\textbf {\bibinfo
  {volume} {100}},\ \bibinfo {pages} {047209} (\bibinfo {year}
  {2008})}\BibitemShut {NoStop}%
\bibitem [{\citenamefont {Son}\ \emph {et~al.}(2006{\natexlab{a}})\citenamefont
  {Son}, \citenamefont {Cohen},\ and\ \citenamefont {Louie}}]{Son2006}%
  \BibitemOpen
  \bibfield  {author} {\bibinfo {author} {\bibfnamefont {Y.-W.}\ \bibnamefont
  {Son}}, \bibinfo {author} {\bibfnamefont {M.~L.}\ \bibnamefont {Cohen}}, \
  and\ \bibinfo {author} {\bibfnamefont {S.~G.}\ \bibnamefont {Louie}},\ }\href
  {\doibase 10.1038/nature05180} {\bibfield  {journal} {\bibinfo  {journal}
  {Nature}\ }\textbf {\bibinfo {volume} {444}},\ \bibinfo {pages} {347}
  (\bibinfo {year} {2006}{\natexlab{a}})}\BibitemShut {NoStop}%
\bibitem [{\citenamefont {Wimmer}\ \emph {et~al.}(2008)\citenamefont {Wimmer},
  \citenamefont {Adagideli}, \citenamefont {Berber}, \citenamefont
  {Tom\'anek},\ and\ \citenamefont {Richter}}]{Wimmer2008}%
  \BibitemOpen
  \bibfield  {author} {\bibinfo {author} {\bibfnamefont {M.}~\bibnamefont
  {Wimmer}}, \bibinfo {author} {\bibfnamefont {I.}~\bibnamefont {Adagideli}},
  \bibinfo {author} {\bibfnamefont {S.}~\bibnamefont {Berber}}, \bibinfo
  {author} {\bibfnamefont {D.}~\bibnamefont {Tom\'anek}}, \ and\ \bibinfo
  {author} {\bibfnamefont {K.}~\bibnamefont {Richter}},\ }\href {\doibase
  10.1103/PhysRevLett.100.177207} {\bibfield  {journal} {\bibinfo  {journal}
  {Phys. Rev. Lett.}\ }\textbf {\bibinfo {volume} {100}},\ \bibinfo {pages}
  {177207} (\bibinfo {year} {2008})}\BibitemShut {NoStop}%
\bibitem [{\citenamefont {Wang}\ \emph {et~al.}(2008)\citenamefont {Wang},
  \citenamefont {Meng},\ and\ \citenamefont {Kaxiras}}]{Wang2008}%
  \BibitemOpen
  \bibfield  {author} {\bibinfo {author} {\bibfnamefont {W.~L.}\ \bibnamefont
  {Wang}}, \bibinfo {author} {\bibfnamefont {S.}~\bibnamefont {Meng}}, \ and\
  \bibinfo {author} {\bibfnamefont {E.}~\bibnamefont {Kaxiras}},\ }\href
  {\doibase 10.1021/nl072548a} {\bibfield  {journal} {\bibinfo  {journal} {Nano
  Lett.}\ }\textbf {\bibinfo {volume} {8}},\ \bibinfo {pages} {241} (\bibinfo
  {year} {2008})}\BibitemShut {NoStop}%
\bibitem [{\citenamefont {Wang}\ \emph {et~al.}(2009)\citenamefont {Wang},
  \citenamefont {Yazyev}, \citenamefont {Meng},\ and\ \citenamefont
  {Kaxiras}}]{Wang2009}%
  \BibitemOpen
  \bibfield  {author} {\bibinfo {author} {\bibfnamefont {W.~L.}\ \bibnamefont
  {Wang}}, \bibinfo {author} {\bibfnamefont {O.~V.}\ \bibnamefont {Yazyev}},
  \bibinfo {author} {\bibfnamefont {S.}~\bibnamefont {Meng}}, \ and\ \bibinfo
  {author} {\bibfnamefont {E.}~\bibnamefont {Kaxiras}},\ }\href {\doibase
  10.1103/PhysRevLett.102.157201} {\bibfield  {journal} {\bibinfo  {journal}
  {Phys. Rev. Lett.}\ }\textbf {\bibinfo {volume} {102}},\ \bibinfo {pages}
  {157201} (\bibinfo {year} {2009})}\BibitemShut {NoStop}%
\bibitem [{\citenamefont {Avsar}\ \emph {et~al.}(2020)\citenamefont {Avsar},
  \citenamefont {Ochoa}, \citenamefont {Guinea}, \citenamefont {\"Ozyilmaz},
  \citenamefont {van Wees},\ and\ \citenamefont {Vera-Marun}}]{Avsar2020}%
  \BibitemOpen
  \bibfield  {author} {\bibinfo {author} {\bibfnamefont {A.}~\bibnamefont
  {Avsar}}, \bibinfo {author} {\bibfnamefont {H.}~\bibnamefont {Ochoa}},
  \bibinfo {author} {\bibfnamefont {F.}~\bibnamefont {Guinea}}, \bibinfo
  {author} {\bibfnamefont {B.}~\bibnamefont {\"Ozyilmaz}}, \bibinfo {author}
  {\bibfnamefont {B.~J.}\ \bibnamefont {van Wees}}, \ and\ \bibinfo {author}
  {\bibfnamefont {I.~J.}\ \bibnamefont {Vera-Marun}},\ }\href@noop {}
  {\bibfield  {journal} {\bibinfo  {journal} {Rev. Mod. Phys.}\ }\textbf
  {\bibinfo {volume} {92}},\ \bibinfo {pages} {021003} (\bibinfo {year}
  {2020})}\BibitemShut {NoStop}%
\bibitem [{\citenamefont {Han}\ \emph {et~al.}(2014)\citenamefont {Han},
  \citenamefont {Kawakami}, \citenamefont {Gmitra},\ and\ \citenamefont
  {Fabian}}]{Han2014}%
  \BibitemOpen
  \bibfield  {author} {\bibinfo {author} {\bibfnamefont {W.}~\bibnamefont
  {Han}}, \bibinfo {author} {\bibfnamefont {R.~K.}\ \bibnamefont {Kawakami}},
  \bibinfo {author} {\bibfnamefont {M.}~\bibnamefont {Gmitra}}, \ and\ \bibinfo
  {author} {\bibfnamefont {J.}~\bibnamefont {Fabian}},\ }\href@noop {}
  {\bibfield  {journal} {\bibinfo  {journal} {Nat. Nanotechnol.}\ }\textbf
  {\bibinfo {volume} {9}},\ \bibinfo {pages} {794} (\bibinfo {year}
  {2014})}\BibitemShut {NoStop}%
\bibitem [{\citenamefont {Chen}\ \emph {et~al.}(2013)\citenamefont {Chen},
  \citenamefont {de~Oteyza}, \citenamefont {Pedramrazi}, \citenamefont {Chen},
  \citenamefont {Fischer},\ and\ \citenamefont {Crommie}}]{Chen13}%
  \BibitemOpen
  \bibfield  {author} {\bibinfo {author} {\bibfnamefont {Y.-C.}\ \bibnamefont
  {Chen}}, \bibinfo {author} {\bibfnamefont {D.~G.}\ \bibnamefont {de~Oteyza}},
  \bibinfo {author} {\bibfnamefont {Z.}~\bibnamefont {Pedramrazi}}, \bibinfo
  {author} {\bibfnamefont {C.}~\bibnamefont {Chen}}, \bibinfo {author}
  {\bibfnamefont {F.~R.}\ \bibnamefont {Fischer}}, \ and\ \bibinfo {author}
  {\bibfnamefont {M.~F.}\ \bibnamefont {Crommie}},\ }\href@noop {} {\bibfield
  {journal} {\bibinfo  {journal} {ACS Nano}\ }\textbf {\bibinfo {volume} {7}},\
  \bibinfo {pages} {6123} (\bibinfo {year} {2013})}\BibitemShut {NoStop}%
\bibitem [{\citenamefont {Son}\ \emph {et~al.}(2006{\natexlab{b}})\citenamefont
  {Son}, \citenamefont {Cohen},\ and\ \citenamefont {Louie}}]{Son06a}%
  \BibitemOpen
  \bibfield  {author} {\bibinfo {author} {\bibfnamefont {Y.-W.}\ \bibnamefont
  {Son}}, \bibinfo {author} {\bibfnamefont {M.~L.}\ \bibnamefont {Cohen}}, \
  and\ \bibinfo {author} {\bibfnamefont {S.~G.}\ \bibnamefont {Louie}},\
  }\href@noop {} {\bibfield  {journal} {\bibinfo  {journal} {Phys. Rev. Lett.}\
  }\textbf {\bibinfo {volume} {97}},\ \bibinfo {pages} {216803} (\bibinfo
  {year} {2006}{\natexlab{b}})}\BibitemShut {NoStop}%
\bibitem [{\citenamefont {Li}\ \emph {et~al.}(2019)\citenamefont {Li},
  \citenamefont {Sanz}, \citenamefont {Corso}, \citenamefont {Choi},
  \citenamefont {Pe{\~{n}}a}, \citenamefont {Frederiksen},\ and\ \citenamefont
  {Pascual}}]{Li2019}%
  \BibitemOpen
  \bibfield  {author} {\bibinfo {author} {\bibfnamefont {J.}~\bibnamefont
  {Li}}, \bibinfo {author} {\bibfnamefont {S.}~\bibnamefont {Sanz}}, \bibinfo
  {author} {\bibfnamefont {M.}~\bibnamefont {Corso}}, \bibinfo {author}
  {\bibfnamefont {D.~J.}\ \bibnamefont {Choi}}, \bibinfo {author}
  {\bibfnamefont {D.}~\bibnamefont {Pe{\~{n}}a}}, \bibinfo {author}
  {\bibfnamefont {T.}~\bibnamefont {Frederiksen}}, \ and\ \bibinfo {author}
  {\bibfnamefont {J.~I.}\ \bibnamefont {Pascual}},\ }\href@noop {} {\bibfield
  {journal} {\bibinfo  {journal} {Nat. Commun.}\ }\textbf {\bibinfo {volume}
  {10}},\ \bibinfo {pages} {200} (\bibinfo {year} {2019})}\BibitemShut
  {NoStop}%
\bibitem [{\citenamefont {Lawrence}\ \emph {et~al.}(2020)\citenamefont
  {Lawrence}, \citenamefont {Brandimarte}, \citenamefont {Berdonces-Layunta},
  \citenamefont {Mohammed}, \citenamefont {Grewal}, \citenamefont {Leon},
  \citenamefont {S{\'a}nchez-Portal},\ and\ \citenamefont
  {de~Oteyza}}]{Lawrence2020}%
  \BibitemOpen
  \bibfield  {author} {\bibinfo {author} {\bibfnamefont {J.}~\bibnamefont
  {Lawrence}}, \bibinfo {author} {\bibfnamefont {P.}~\bibnamefont
  {Brandimarte}}, \bibinfo {author} {\bibfnamefont {A.}~\bibnamefont
  {Berdonces-Layunta}}, \bibinfo {author} {\bibfnamefont {M.~S.~G.}\
  \bibnamefont {Mohammed}}, \bibinfo {author} {\bibfnamefont {A.}~\bibnamefont
  {Grewal}}, \bibinfo {author} {\bibfnamefont {C.~C.}\ \bibnamefont {Leon}},
  \bibinfo {author} {\bibfnamefont {D.}~\bibnamefont {S{\'a}nchez-Portal}}, \
  and\ \bibinfo {author} {\bibfnamefont {D.~G.}\ \bibnamefont {de~Oteyza}},\
  }\href {\doibase 10.1021/acsnano.9b10191} {\bibfield  {journal} {\bibinfo
  {journal} {ACS Nano}\ }\textbf {\bibinfo {volume} {14}},\ \bibinfo {pages}
  {4499} (\bibinfo {year} {2020})}\BibitemShut {NoStop}%
\bibitem [{\citenamefont {Yazyev}(2010)}]{Yazyev2010}%
  \BibitemOpen
  \bibfield  {author} {\bibinfo {author} {\bibfnamefont {O.~V.}\ \bibnamefont
  {Yazyev}},\ }\href@noop {} {\bibfield  {journal} {\bibinfo  {journal} {Rep.
  Prog. Phys.}\ }\textbf {\bibinfo {volume} {73}},\ \bibinfo {pages} {056501}
  (\bibinfo {year} {2010})}\BibitemShut {NoStop}%
\bibitem [{\citenamefont {Tu\v{c}ek}\ \emph {et~al.}(2018)\citenamefont
  {Tu\v{c}ek}, \citenamefont {B\l'o\'{n}ski}, \citenamefont {Ugolotti},
  \citenamefont {Swain}, \citenamefont {Enoki},\ and\ \citenamefont
  {Zbořil}}]{Radek18}%
  \BibitemOpen
  \bibfield  {author} {\bibinfo {author} {\bibfnamefont {J.}~\bibnamefont
  {Tu\v{c}ek}}, \bibinfo {author} {\bibfnamefont {P.}~\bibnamefont
  {B\l'o\'{n}ski}}, \bibinfo {author} {\bibfnamefont {J.}~\bibnamefont
  {Ugolotti}}, \bibinfo {author} {\bibfnamefont {A.~K.}\ \bibnamefont {Swain}},
  \bibinfo {author} {\bibfnamefont {T.}~\bibnamefont {Enoki}}, \ and\ \bibinfo
  {author} {\bibfnamefont {R.}~\bibnamefont {Zbořil}},\ }\href@noop {}
  {\bibfield  {journal} {\bibinfo  {journal} {Chem. Soc. Rev.}\ }\textbf
  {\bibinfo {volume} {47}},\ \bibinfo {pages} {3899} (\bibinfo {year}
  {2018})}\BibitemShut {NoStop}%
\bibitem [{\citenamefont {Sun}\ \emph {et~al.}(2020{\natexlab{b}})\citenamefont
  {Sun}, \citenamefont {Yao}, \citenamefont {Gr{\"o}ning}, \citenamefont {Eimre},
  \citenamefont {Pignedoli}, \citenamefont {M{\"u}llen}, \citenamefont {Narita},
  \citenamefont {Fasel},\ and\ \citenamefont {Ruffieux}}]{Sun2020m}%
  \BibitemOpen
  \bibfield  {author} {\bibinfo {author} {\bibfnamefont {Q.}~\bibnamefont
  {Sun}}, \bibinfo {author} {\bibfnamefont {X.}~\bibnamefont {Yao}}, \bibinfo
  {author} {\bibfnamefont {O.}~\bibnamefont {Gr{\"o}ning}}, \bibinfo {author}
  {\bibfnamefont {K.}~\bibnamefont {Eimre}}, \bibinfo {author} {\bibfnamefont
  {C.~A.}\ \bibnamefont {Pignedoli}}, \bibinfo {author} {\bibfnamefont
  {K.}~\bibnamefont {M\"ullen}}, \bibinfo {author} {\bibfnamefont
  {A.}~\bibnamefont {Narita}}, \bibinfo {author} {\bibfnamefont
  {R.}~\bibnamefont {Fasel}}, \ and\ \bibinfo {author} {\bibfnamefont
  {P.}~\bibnamefont {Ruffieux}},\ }\href@noop {} {\bibfield  {journal}
  {\bibinfo  {journal} {Nano Lett.}\ }\textbf {\bibinfo {volume} {20}},\
  \bibinfo {pages} {6429} (\bibinfo {year} {2020}{\natexlab{b}})}\BibitemShut
  {NoStop}%
\bibitem [{\citenamefont {Rizzo}\ \emph {et~al.}(2020)\citenamefont {Rizzo},
  \citenamefont {Veber}, \citenamefont {Jiang}, \citenamefont {McCurdy},
  \citenamefont {Cao}, \citenamefont {Bronner}, \citenamefont {Chen},
  \citenamefont {Louie}, \citenamefont {Fischer},\ and\ \citenamefont
  {Crommie}}]{Rizzo1597}%
  \BibitemOpen
  \bibfield  {author} {\bibinfo {author} {\bibfnamefont {D.~J.}\ \bibnamefont
  {Rizzo}}, \bibinfo {author} {\bibfnamefont {G.}~\bibnamefont {Veber}},
  \bibinfo {author} {\bibfnamefont {J.}~\bibnamefont {Jiang}}, \bibinfo
  {author} {\bibfnamefont {R.}~\bibnamefont {McCurdy}}, \bibinfo {author}
  {\bibfnamefont {T.}~\bibnamefont {Cao}}, \bibinfo {author} {\bibfnamefont
  {C.}~\bibnamefont {Bronner}}, \bibinfo {author} {\bibfnamefont
  {T.}~\bibnamefont {Chen}}, \bibinfo {author} {\bibfnamefont {S.~G.}\
  \bibnamefont {Louie}}, \bibinfo {author} {\bibfnamefont {F.~R.}\ \bibnamefont
  {Fischer}}, \ and\ \bibinfo {author} {\bibfnamefont {M.~F.}\ \bibnamefont
  {Crommie}},\ }\href {\doibase 10.1126/science.aay3588} {\bibfield  {journal}
  {\bibinfo  {journal} {Science}\ }\textbf {\bibinfo {volume} {369}},\ \bibinfo
  {pages} {1597} (\bibinfo {year} {2020})}\BibitemShut {NoStop}%
\bibitem [{\citenamefont {Perdew}\ \emph {et~al.}(1996)\citenamefont {Perdew},
  \citenamefont {Burke},\ and\ \citenamefont {Ernzerhof}}]{PBE}%
  \BibitemOpen
  \bibfield  {author} {\bibinfo {author} {\bibfnamefont {J.~P.}\ \bibnamefont
  {Perdew}}, \bibinfo {author} {\bibfnamefont {K.}~\bibnamefont {Burke}}, \
  and\ \bibinfo {author} {\bibfnamefont {M.}~\bibnamefont {Ernzerhof}},\ }\href
  {\doibase 10.1103/PhysRevLett.77.3865} {\bibfield  {journal} {\bibinfo
  {journal} {Phys. Rev. Lett.}\ }\textbf {\bibinfo {volume} {77}},\ \bibinfo
  {pages} {3865} (\bibinfo {year} {1996})}\BibitemShut {NoStop}%
\bibitem [{\citenamefont {Soler}\ \emph {et~al.}(2002)\citenamefont {Soler},
  \citenamefont {Artacho}, \citenamefont {Gale}, \citenamefont {Garc{\'{\i}}a},
  \citenamefont {Junquera}, \citenamefont {Ordej{\'{o}}n},\ and\ \citenamefont
  {S{\'a}nchez-Portal}}]{SIESTA}%
  \BibitemOpen
  \bibfield  {author} {\bibinfo {author} {\bibfnamefont {J.~M.}\ \bibnamefont
  {Soler}}, \bibinfo {author} {\bibfnamefont {E.}~\bibnamefont {Artacho}},
  \bibinfo {author} {\bibfnamefont {J.~D.}\ \bibnamefont {Gale}}, \bibinfo
  {author} {\bibfnamefont {A.}~\bibnamefont {Garc{\'{\i}}a}}, \bibinfo {author}
  {\bibfnamefont {J.}~\bibnamefont {Junquera}}, \bibinfo {author}
  {\bibfnamefont {P.}~\bibnamefont {Ordej{\'{o}}n}}, \ and\ \bibinfo {author}
  {\bibfnamefont {D.}~\bibnamefont {S{\'a}nchez-Portal}},\ }\href@noop {}
  {\bibfield  {journal} {\bibinfo  {journal} {J. Phys. Condens. Matter}\
  }\textbf {\bibinfo {volume} {14}},\ \bibinfo {pages} {2745} (\bibinfo {year}
  {2002})}\BibitemShut {NoStop}%
\bibitem [{\citenamefont {Papior}\ \emph {et~al.}(2017)\citenamefont {Papior},
  \citenamefont {Lorente}, \citenamefont {Frederiksen}, \citenamefont
  {Garc\'i�a},\ and\ \citenamefont {Brandbyge}}]{TRANSIESTA}%
  \BibitemOpen
  \bibfield  {author} {\bibinfo {author} {\bibfnamefont {N.}~\bibnamefont
  {Papior}}, \bibinfo {author} {\bibfnamefont {N.}~\bibnamefont {Lorente}},
  \bibinfo {author} {\bibfnamefont {T.}~\bibnamefont {Frederiksen}}, \bibinfo
  {author} {\bibfnamefont {A.}~\bibnamefont {Garc\'i�a}}, \ and\ \bibinfo
  {author} {\bibfnamefont {M.}~\bibnamefont {Brandbyge}},\ }\href@noop {}
  {\bibfield  {journal} {\bibinfo  {journal} {Comput. Phys. Commun.}\ }\textbf
  {\bibinfo {volume} {212}},\ \bibinfo {pages} {8} (\bibinfo {year}
  {2017})}\BibitemShut {NoStop}%
\bibitem [{\citenamefont {Lieb}(1989)}]{Lieb1989}%
  \BibitemOpen
  \bibfield  {author} {\bibinfo {author} {\bibfnamefont {E.~H.}\ \bibnamefont
  {Lieb}},\ }\href {\doibase 10.1103/PhysRevLett.62.1201} {\bibfield  {journal}
  {\bibinfo  {journal} {Phys. Rev. Lett.}\ }\textbf {\bibinfo {volume} {62}},\
  \bibinfo {pages} {1201} (\bibinfo {year} {1989})}\BibitemShut {NoStop}%
\bibitem [{\citenamefont {Hirjibehedin}\ \emph {et~al.}(2006)\citenamefont
  {Hirjibehedin}, \citenamefont {Lutz},\ and\ \citenamefont
  {Heinrich}}]{Hirjibehedin1021}%
  \BibitemOpen
  \bibfield  {author} {\bibinfo {author} {\bibfnamefont {C.~F.}\ \bibnamefont
  {Hirjibehedin}}, \bibinfo {author} {\bibfnamefont {C.~P.}\ \bibnamefont
  {Lutz}}, \ and\ \bibinfo {author} {\bibfnamefont {A.~J.}\ \bibnamefont
  {Heinrich}},\ }\href {\doibase 10.1126/science.1125398} {\bibfield  {journal}
  {\bibinfo  {journal} {Science}\ }\textbf {\bibinfo {volume} {312}},\ \bibinfo
  {pages} {1021} (\bibinfo {year} {2006})}\BibitemShut {NoStop}%
\bibitem [{\citenamefont {Jung}\ and\ \citenamefont
  {MacDonald}(2010)}]{Jung2010}%
  \BibitemOpen
  \bibfield  {author} {\bibinfo {author} {\bibfnamefont {J.}~\bibnamefont
  {Jung}}\ and\ \bibinfo {author} {\bibfnamefont {A.~H.}\ \bibnamefont
  {MacDonald}},\ }\href {\doibase 10.1103/PhysRevB.81.195408} {\bibfield
  {journal} {\bibinfo  {journal} {Phys. Rev. B}\ }\textbf {\bibinfo {volume}
  {81}},\ \bibinfo {pages} {195408} (\bibinfo {year} {2010})}\BibitemShut
  {NoStop}%
\bibitem [{\citenamefont {Pizzochero}\ \emph {et~al.}(2015)\citenamefont
  {Pizzochero}, \citenamefont {Leenaerts}, \citenamefont {Partoens},
  \citenamefont {Martinazzo},\ and\ \citenamefont {Peeters}}]{Pizzochero2015}%
  \BibitemOpen
  \bibfield  {author} {\bibinfo {author} {\bibfnamefont {M.}~\bibnamefont
  {Pizzochero}}, \bibinfo {author} {\bibfnamefont {O.}~\bibnamefont
  {Leenaerts}}, \bibinfo {author} {\bibfnamefont {B.}~\bibnamefont {Partoens}},
  \bibinfo {author} {\bibfnamefont {R.}~\bibnamefont {Martinazzo}}, \ and\
  \bibinfo {author} {\bibfnamefont {F.~M.}\ \bibnamefont {Peeters}},\ }\href
  {\doibase 10.1088/0953-8984/27/42/425502} {\bibfield  {journal} {\bibinfo
  {journal} {J. Phys. Condens. Matter}\ }\textbf {\bibinfo {volume} {27}},\
  \bibinfo {pages} {425502} (\bibinfo {year} {2015})}\BibitemShut {NoStop}%
\bibitem [{\citenamefont {Nair}\ \emph {et~al.}(2013)\citenamefont {Nair},
  \citenamefont {Tsai}, \citenamefont {Sepioni}, \citenamefont {Lehtinen},
  \citenamefont {Keinonen}, \citenamefont {Krasheninnikov}, \citenamefont
  {Castro~Neto}, \citenamefont {Katsnelson}, \citenamefont {Geim},\ and\
  \citenamefont {Grigorieva}}]{Nair2013}%
  \BibitemOpen
  \bibfield  {author} {\bibinfo {author} {\bibfnamefont {R.~R.}\ \bibnamefont
  {Nair}}, \bibinfo {author} {\bibfnamefont {I.-L.}\ \bibnamefont {Tsai}},
  \bibinfo {author} {\bibfnamefont {M.}~\bibnamefont {Sepioni}}, \bibinfo
  {author} {\bibfnamefont {O.}~\bibnamefont {Lehtinen}}, \bibinfo {author}
  {\bibfnamefont {J.}~\bibnamefont {Keinonen}}, \bibinfo {author}
  {\bibfnamefont {A.~V.}\ \bibnamefont {Krasheninnikov}}, \bibinfo {author}
  {\bibfnamefont {A.~H.}\ \bibnamefont {Castro~Neto}}, \bibinfo {author}
  {\bibfnamefont {M.~I.}\ \bibnamefont {Katsnelson}}, \bibinfo {author}
  {\bibfnamefont {A.~K.}\ \bibnamefont {Geim}}, \ and\ \bibinfo {author}
  {\bibfnamefont {I.~V.}\ \bibnamefont {Grigorieva}},\ }\href@noop {}
  {\bibfield  {journal} {\bibinfo  {journal} {Nat. Commun.}\ }\textbf {\bibinfo
  {volume} {4}},\ \bibinfo {pages} {2010} (\bibinfo {year} {2013})}\BibitemShut
  {NoStop}%
\bibitem [{\citenamefont {Landauer}(1961)}]{Landauer}%
  \BibitemOpen
  \bibfield  {author} {\bibinfo {author} {\bibfnamefont {R.}~\bibnamefont
  {Landauer}},\ }\href@noop {} {\bibfield  {journal} {\bibinfo  {journal} {IBM
  J. Res. Dev.}\ }\textbf {\bibinfo {volume} {5}},\ \bibinfo {pages} {183}
  (\bibinfo {year} {1961})}\BibitemShut {NoStop}%
\end{thebibliography}
 \end{document}